\documentclass[preprint,3p,twocolumn,sort&compress]{elsarticle}
\usepackage{float}
\usepackage{soul}
\usepackage{amsmath}
\usepackage{graphicx}
\usepackage{dcolumn}
\usepackage{bm}
\usepackage{natbib}
\usepackage{tikz}
\usepackage{color}
\usepackage{subfigure}
\usepackage{longtable}
\usepackage{dcolumn}
\usepackage{tabularx,multirow}
\usepackage[colorlinks, linkcolor={blue},citecolor={blue},breaklinks]{hyperref}
\journal{Phys. Lett. B}

\begin{document}

\begin{frontmatter}

\title{Low-energy $^{23}$Al $\beta$-delayed proton decay and $^{22}$Na destruction in novae}
\author[NSCLaddress,HUJIaddress]{M. Friedman\corref{mycorrespondingauthor}}
\cortext[mycorrespondingauthor]{Corresponding author}
\ead{moshe.friedman@mail.huji.ac.il}
\author[NSCLaddress,MSUaddress]{T. Budner}
\author[NSCLaddress,UTennesse]{D. P\'erez-Loureiro}
\author[Saclay]{E. Pollacco}
\author[NSCLaddress,MSUaddress]{C. Wrede}
\author[UPC,IEEC]{J. Jos{\'e}}
\author[NSCLaddress,MSUaddress]{B. A. Brown}
\author[NSCLaddress]{M. Cortesi}
\author[NSCLaddress,MSUaddress]{C. Fry}
\author[NSCLaddress,MSUaddress]{B. Glassman}
\author[UTennesse]{J. Heideman}
\author[NSCLaddress,MSUaddress]{M. Janasik}
\author[NSCLaddress,MSUaddress]{M. Roosa}
\author[NSCLaddress,MSUaddress]{J. Stomps}
\author[NSCLaddress,MSUaddress]{J. Surbrook}
\author[NSCLaddress,MSUaddress]{P. Tiwari}

\address[NSCLaddress]{National Superconducting Cyclotron Laboratory, Michigan State University, East Lansing, Michigan 48824, USA}
\address[HUJIaddress]{Racah Institute of Physics, Hebrew University of Jerusalem, Jerusalem, Israel 91904}
\address[MSUaddress]{Department of Physics and Astronomy,  Michigan State University, East Lansing, Michigan 48824, USA}
\address[UTennesse]{Department of Physics and Astronomy,  University of Tennessee, Knoxville, Tennessee , 37996 USA}
\address[Saclay]{IRFU, CEA, Universit\'e Paris-Saclay, F-91191, Gif-sur-Yvette, France}
\address[UPC]{Departament de F{\'i}sica, Escola d'Enginyeria de Barcelona Est, Universitat Polit{\`e}cnica de Catalunya, Av./ Eduard Maristany 10, E-08930 Barcelona, Spain}
\address[IEEC]{Institut d'Estudis Espacials de Catalunya, Ed. Nexus-201, C/ Gran Capit{\`a} 2-4, E-08034 Barcelona, Spain}

\begin{abstract}
The radionuclide $^{22}$Na is a target of $\gamma$-ray astronomy searches, predicted to be produced during thermonuclear runaways driving classical novae. The $^{22}$Na(p,$\gamma$)$^{23}$Mg reaction is the main destruction channel of $^{22}$Na during a nova, hence,  its rate is needed to accurately predict the $^{22}$Na yield. However, experimental determinations of the resonance strengths have led to inconsistent results. In this work, we report a measurement of the branching ratios of the $^{23}$Al $\beta$-delayed protons, as a probe of the key 204--keV (center-of-mass) $^{22}$Na(p,$\gamma$)$^{23}$Mg resonance strength. We report a factor of 5 lower branching ratio compared to the most recent literature value. The variation in $^{22}$Na yield due to nuclear data inconsistencies was assessed using a series of hydrodynamic nova outburst simulations and has increased to a factor of 3.8, corresponding to a factor of $\sim$2 uncertainty in the maximum detectability distance. This is the first reported scientific measurement using the Gaseous Detector with Germanium Tagging (GADGET) system.
\end{abstract}


\end{frontmatter}

\section{Introduction}

Radionuclides are now routinely observed astronomically using space-based gamma-ray observatories capable of detecting their characteristic emission lines. For example, $^{26}$Al ($T_{1/2} = 0.72$~Ma) and $^{60}$Fe ($T_{1/2} = 2.6$~Ma) live long enough to migrate from the stellar events producing them before they decay and manifest as diffuse emission across the Milky Way. Attempts have been made to benchmark nucleosynthesis in massive stars and their supernovae using the relative amounts of observed $^{26}$Al \cite{ma84apj,sh85apj,ma87apj,di95aa,di06nat} and $^{60}$Fe \cite{wa07aa}, but such benchmarks can only be applied by considering all possible sources in aggregate. More stringent constraints on astrophysical models of particular events can be derived using shorter-lived nuclides that manifest as localized sources such as $^{44}$Ti ($T_{1/2} = 59$~a), which has been observed in the 350-year-old Cas A core-collapse supernova remnant \cite{iy94aa,vi01apj,ro03apj,re06apj,gr18nat} and the younger remnant of Supernova 1987A \cite{bo15sci}. Similarly, the detection of $^{22}$Na ($T_{1/2} = 2.6$~a) from a classical nova explosion has been a long sought constraint \cite{cl74apj,we90aa,st93pr,co95aa,po95apj,Gomez1998,jo99apj}. Observations with previously and currently deployed instruments may have been on the cusp of detecting $^{22}$Na \cite{hi87apj,iy95aas,wa99apj,je00mnr,Starrfield2016,si18}, and more sensitive future missions are being planned \cite{DeAngelis2018,fryer2019catching}. Accurate model predictions of $^{22}$Na nucleosynthesis in novae are needed to estimate the detectability distance and for comparison to past searches and future observations. In addition, $^{20}$Ne/$^{22}$Ne ratios in presolar grains may be used to identify presolar nova grains \cite{Amari2001,Amari2002,Jose2004,Bose2019}. Usually, neon is incorporated in grains via implantation, since noble gases do not condense as stable compounds into grains. However, low $^{20}$Ne/$^{22}$Ne ratios suggests that $^{22}$Ne originated from \textit{in situ} decay of $^{22}$Na.

The $^{22}$Na yield predicted by nova models is sensitive to the thermonuclear rates of the reactions associated with explosive hydrogen burning on the surface of a white-dwarf star accreting hydrogen-rich material from a binary companion star \cite{jo99apj,il02apj}. This has motivated the development of innovative experimental nuclear physics techniques, providing rates to improve predictions of the $^{22}$Na yield \cite{Gorres1982,se90npa,st96npa,sc97prl,bi03prl,Savard2004,Mukherjee2004,Dauria2004,je04prl,Parikh2005,sa10prl,sa11prc,saa11prc,Lovely2019}. While $^{22}$Na is being produced during the thermonuclear runaway driving a nova, the $^{22}$Na(p,$\gamma$)$^{23}$Mg reaction is actively destroying it. The $^{22}$Na yield is related inversely to the reaction rate and, in particular, to the strength of a single resonance at a center-of-mass energy of 204 keV\footnote{This resonance is referred to by the 213--keV proton lab energy in some publications \cite{st96npa,sa11prc}.}. Two direct measurements (\cite{se90npa,st96npa} and \cite{sa10prl,sa11prc}) of the resonance strength using proton beams and radioactive $^{22}$Na targets have yielded values that differ by a factor of 3.2, which results in a factor of $\approx 2$ variation in the expected $^{22}$Na yield from classical novae \cite{sa10prl}. Another way to determine the strength is to combine measurements of the proton branching ratio $\Gamma_p/\Gamma$ of the resonance with its lifetime $\tau$ and spin. The most precise literature values \cite{je04prl,saa11prc} for these quantities yield a strength that is consistent with that from Refs. \cite{se90npa,st96npa}. It may be tempting to consider this the conclusive arbiter for the two inconsistent directly-measured values, but it would be prudent to confirm the proton branching ratio and lifetime, as the aforementioned experiments faced significant challenges.

In particular, the $^{23}$Al$(\beta p)$ experiment of Ref. \cite{saa11prc} suffered from overwhelming $\beta$ background and relied heavily on a background subtraction model. Pollacco \emph{et al.} \cite{Pollacco2013} showed that a gas filled detector can be used to overcome this problem, and  found indications of a branching ratio for the 204--keV level significantly lower than in Ref. \cite{saa11prc}.  Presently, we report a new proton branching ratio for the 204--keV resonance, determined based on measurements of low-energy $^{23}$Al $\beta$-delayed protons with a new system: the Gaseous Detector with Germanium Tagging (GADGET) \cite{Friedman2019}. GADGET is optimized for the detection of low-energy, low-intensity $\beta$-delayed protons with complementary high-resolution high-efficiency $\gamma$-ray detection. This Letter reports the first scientific results from GADGET.

\section{Experiment Setup}

The experiment was performed at the National Superconducting Cyclotron Laboratory (NSCL), where a radioactive  beam of $^{23}$Al was produced by projectile fragmentation. A 150--MeV/u, 75--pnA primary beam of $^{36}$Ar was accelerated using the Coupled Cyclotron Facility \cite{Marti2001} and impinged upon a $^9$Be transmission target, 1363 mg/cm$^2$ in thickness.  $^{23}$Al was isolated in flight to a purity of 69 \% using the A1900 magnetic fragment separator \cite{Stolz2005} incorporating a 300 mg/cm$^2$ Al wedge and the Radio Frequency Fragment Separator (RFFS) \cite{RFFS}.  Upon exiting the RFFS, situated about 6 m upstream of GADGET, the main contaminants were $^{21}$Na, $^{22}$Mg, and $^{16}$N, in decreasing order of intensity, as identified with standard $\Delta E$-ToF method. The beam rate was about 2000 $^{23}$Al ions per second. To optimize the $^{23}$Al beam energy and range, a 2--mm thick rotatable aluminum degrader, located directly in front of the detection system, was used.

\begin{figure}
	\centering
	\includegraphics[width=0.4\textwidth]{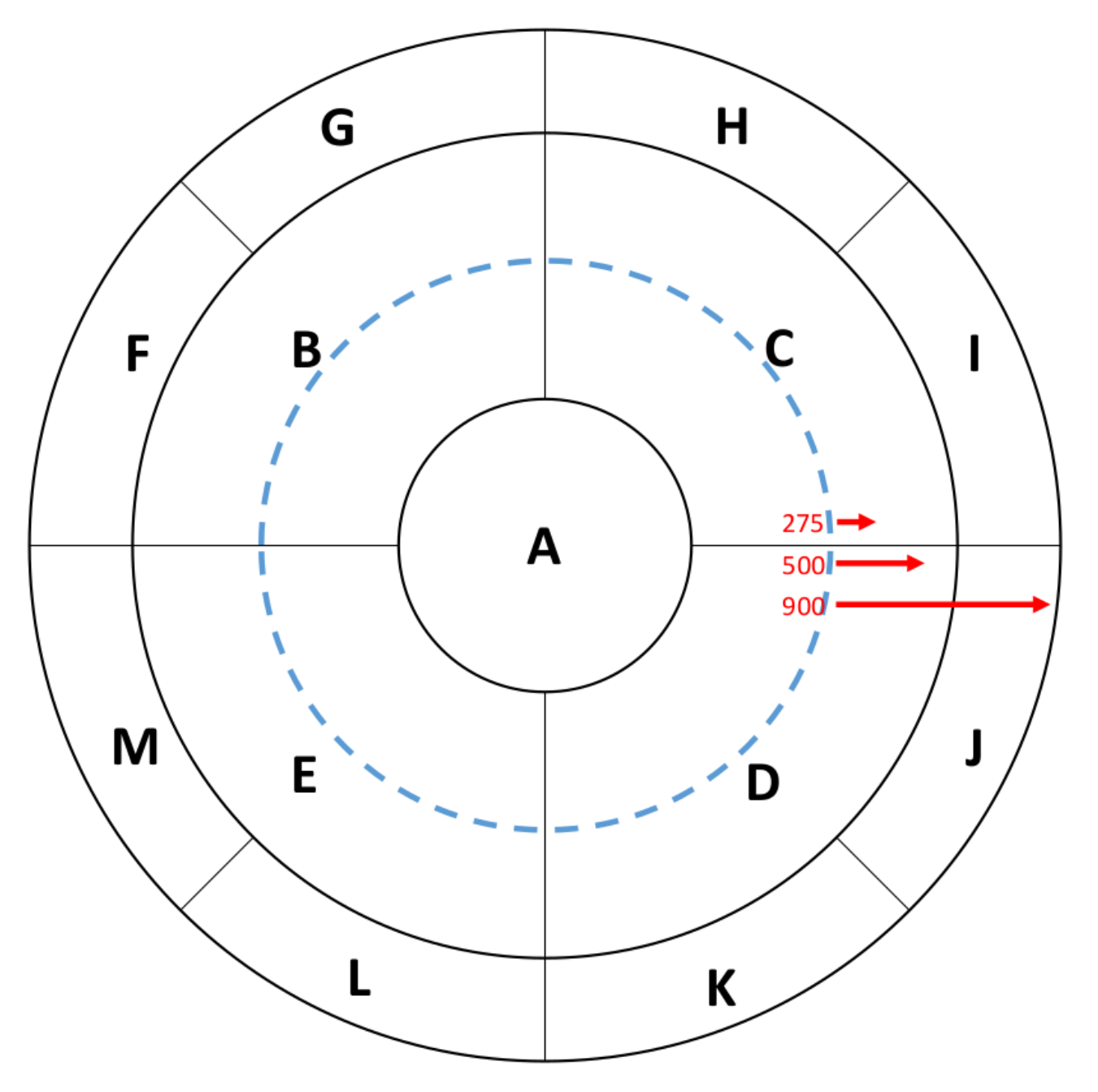}
	\caption{Pad geometry of the anode plane of GADGET's Proton Detector. The radii of the circular borders are 1.4, 4 and 5 cm. An aperture on the cathode limits the transverse dispersion of the beam spot within $\sim$2.6 cm from the center (dashed circle). Ranges of 275--, 500-- and 900--keV protons are illustrated by red arrows.}
	\label{fig:padplane}
\end{figure}

A detailed description of GADGET can be found in \cite{Friedman2019}. Briefly, the assembly contains the Proton Detector, which is a cylindrical vessel filled with P10 gas (set to 780 Torr for this experiment) that functions both as a beam stop and a charged-particle detection medium, surrounded by the Segmented Germanium Array (SeGA) \cite{Mueller2001} for coincident $\gamma$--ray detection. The beam was operated in a pulsed mode, where $^{23}$Al ions ($T_{1/2}=470$ ms) were accumulated in the Proton Detector for 0.5 s, and then the beam was stopped for another 0.5 s to allow their charged particle decay radiations to be detected by ionization. The ionization electrons were drifted towards the readout plane by an uniform electric field of 125 V/cm and amplified by a MICROMEGAS structure \cite{Giomataris2006}. An electrostatic gating grid was used to protect the MICROMEGAS from the large signals produced during the implantation periods. The active volume is a cylinder, 40 cm long and 10 cm in diameter. The short range of the protons in the gas versus the near-transparency to the $\beta$--particles enables the detection of the weak, low--energy, protons and effectively eliminates an otherwise overwhelming $\beta$ background. The MICROMEGAS pad plane is divided into 13 pads, labeled A-M as shown in Fig. \ref{fig:padplane}. This configuration allows vetoing of high--energy protons that might escape the active volume and deposit only part of their energy in the active volume. In addition, it enables analysis of either most of the full active volume (pads A-E) for higher efficiency, or to limit the active volume into sub-sections to achieve lower $\beta$ background at the cost of efficiency loss for higher energy protons. In the current experiment, pads F,G,L,M were not instrumented which resulted in 50\% veto efficiency for events in the A-E active volume. However, in the case of $^{23}$Al only a small fraction of the protons are emitted above $\sim$1 MeV \cite{Kirsebom2011}, hence the associated background is insignificant.

\begin{figure}
   \centering
   \includegraphics[width=0.5\textwidth]{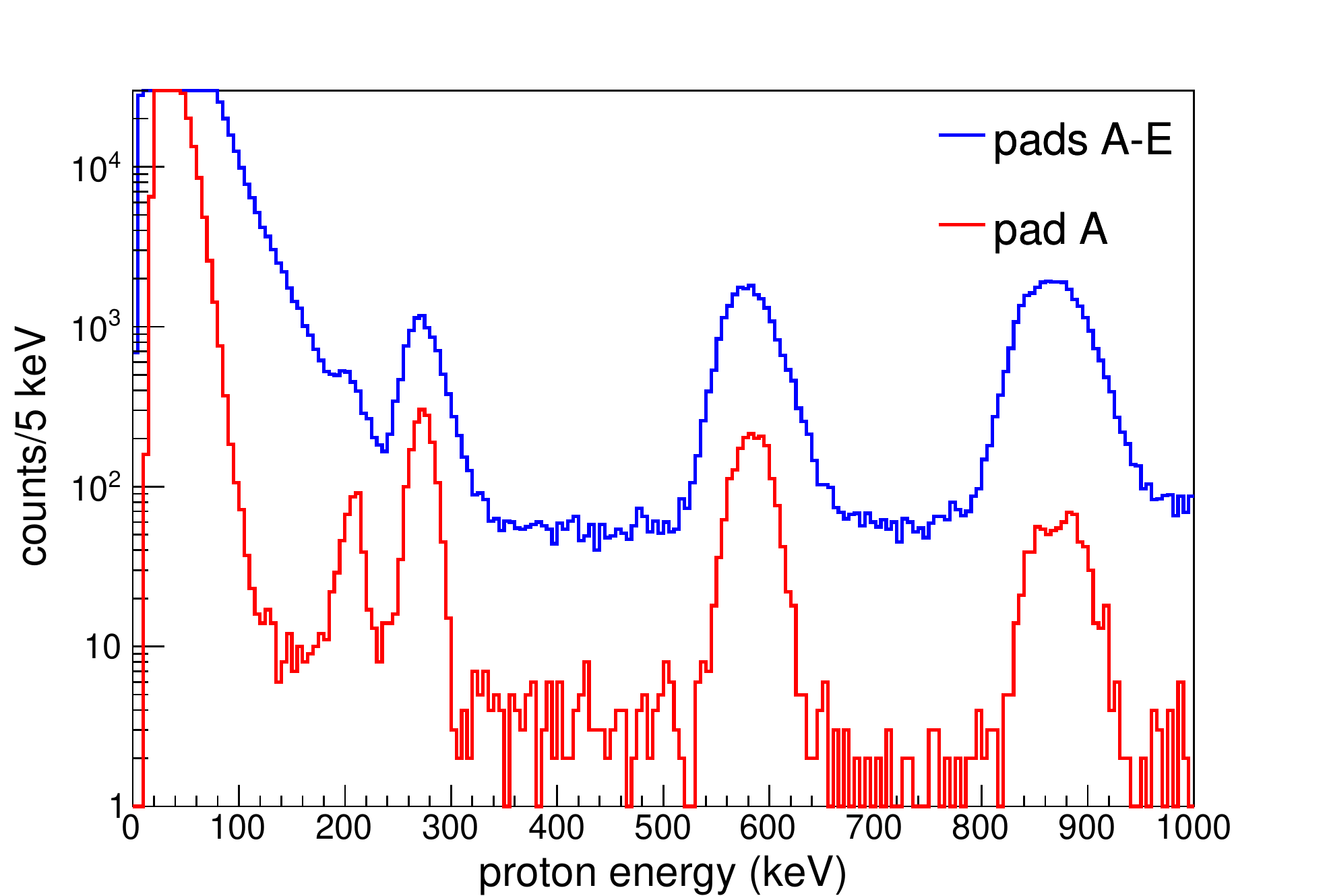}
   \caption{Proton spectrum for pads A-E (blue) and for the central pad alone (red).}
   \label{fig:spectrum}
\end{figure}

 \section{Data Analysis}
 \label{sec:analysis}

\begin{figure}
	\centering
	\includegraphics[width=0.4\textwidth]{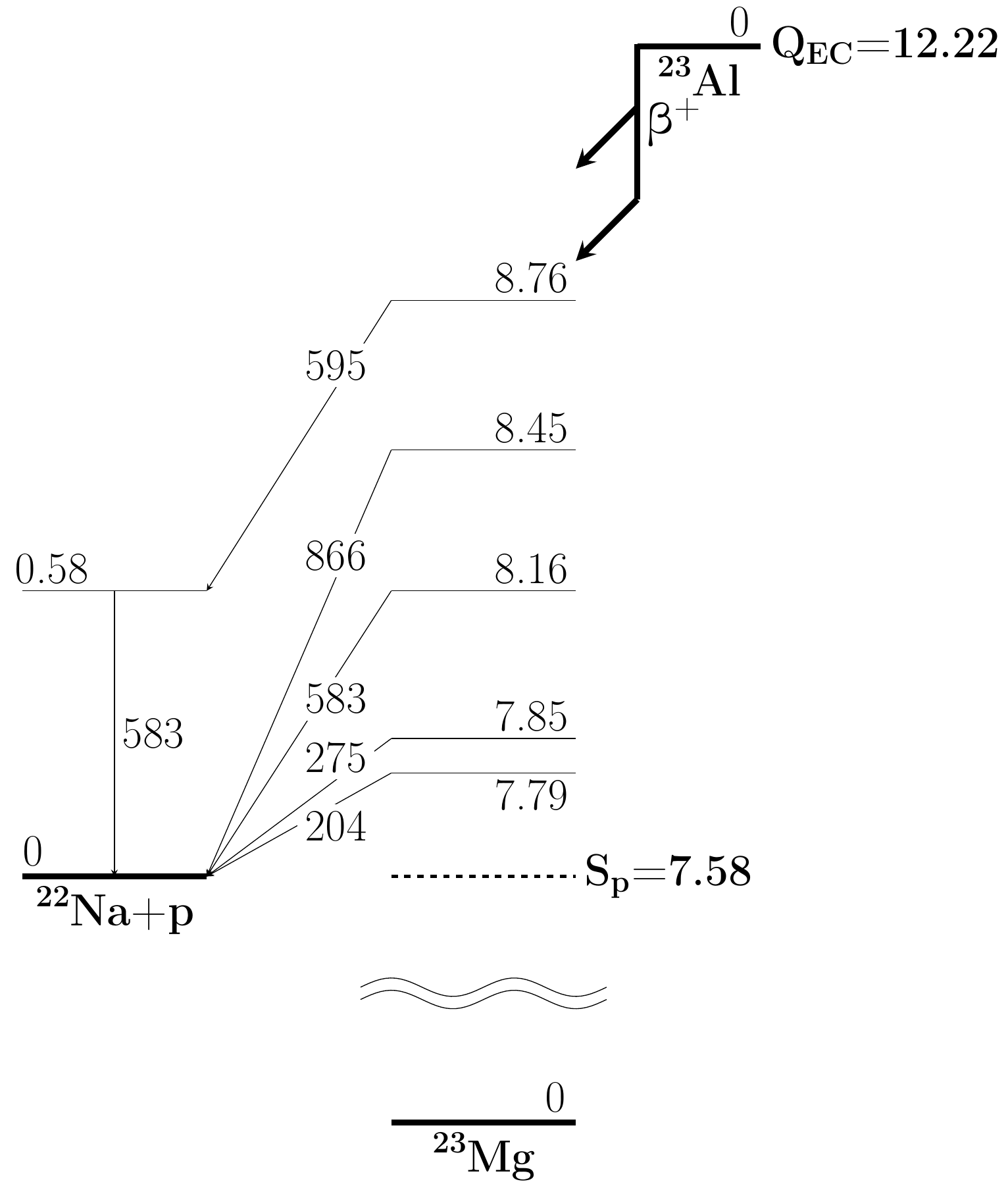}
	\caption{A simplified decay scheme for $^{23}$Al. The scheme only shows the protons with energies up to 900 keV, and an associated $\gamma$-ray. Level  energies are given in MeV and adopted from NuDat \cite{Firestone2007}, while proton (center-of-mass) energies are given in keV and adopted from \cite{sa10prl} and \cite{saa11prc}. See Sec. \ref{sec:analysis} for details.}
	\label{fig:decay_scheme}
\end{figure}

\begin{figure}
   \centering
   \includegraphics[width=0.5\textwidth]{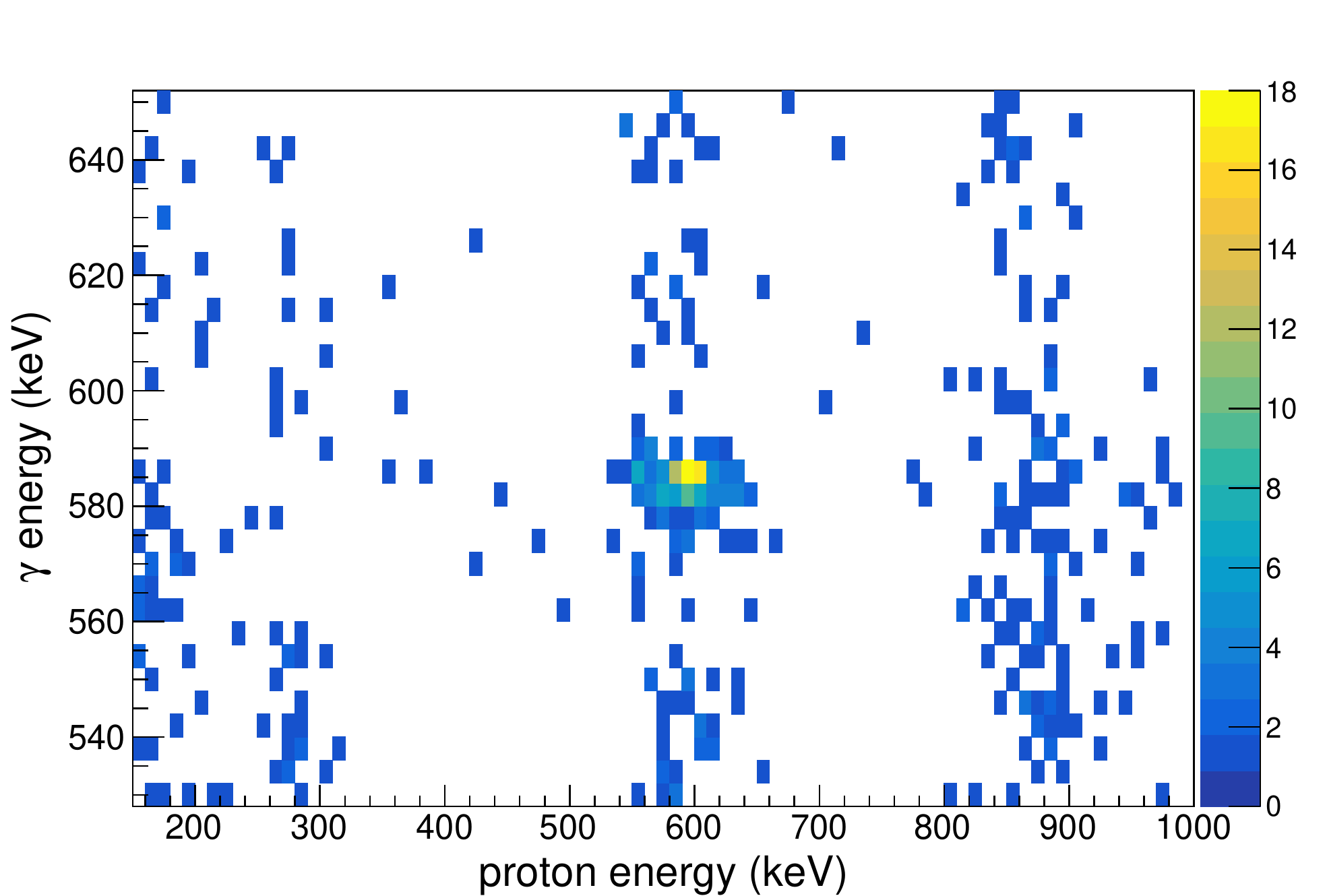}
   \caption{Coincidence spectrum for pads A-E. While no coincidence for the 204--, 275-- or 866--keV protons is seen in the spectrum, a coincidence between 583--keV $\gamma$--rays and 595--keV protons can be seen.}
   \label{fig:coincidence}
\end{figure}

 Fig. \ref{fig:spectrum} shows the proton energy spectrum for pad A (red) and for event-level summing of the five central pads (A--E, blue). The $\beta$ background is substantially suppressed relative to previous experiments \cite{ti95prc,saa11prc}, allowing the extraction of the 204--keV peak intensity from both spectra. The single pad spectrum shows further reduced $\beta$ background due to the smaller active volume, at the cost of a fast-declining efficiency as function of proton energy. There are no beam contaminants that emit $\beta$--delayed charged particles, with the exception of the $^{16}$N that has small $\alpha$--particle branching ratio ($1.2\times10^{-5}$). In addition, those emissions are mostly in an energy range of 1.5 MeV to 3 MeV \cite{France1997,Tang2010,Kirsebom2018}, which is significantly above the region of interest for this experiment. For interpretation of the proton spectrum a simplified decay scheme is presented in Fig. \ref{fig:decay_scheme}. The energy calibration of the spectrum is internal and based on the proton peak energies reported by Ref. \cite{sa10prl} for values up to 583 keV, and on Ref. \cite{saa11prc} for the 866--keV proton peak energy. The spectrum shows linearity with the reported proton energies. In Fig. \ref{fig:decay_scheme} we report a transition from the 8.76--MeV state in $^{23}$Mg to the first excited state of $^{22}$Na, corresponding to a 595--keV proton emission. This transition is identified based on $\gamma-p$ coincidences, as shown in Fig. \ref{fig:coincidence}. The detector resolution is not sufficient to separate this proton peak from the 583--keV proton peak. However, the intensity of the coincidences is too low to account for the full intensity of the 583--keV proton peak, and the centroid of the correlated proton peak is at 595 keV. Therefore, we interpret this proton peak as a double peak with two similar energies. The corrected decay scheme should affect the interpretation of the proton intensities reported by Refs. \cite{ti95prc,pe00plb,saa11prc,Sun2015}. We do not find the 204--keV protons in coincidence with any $\gamma$--rays except for annihilation 511--keV $\gamma$--rays (see Fig. \ref{fig:coincidence}). This confirms that the final state of the proton emission is the ground state of $^{22}$Na, and therefore, the initial state is the resonance of interest.

The intensities of the 204--, 275-- and 583--keV peaks were normalized relative to the 866--keV peak. Although the detection efficiency of the low-energy protons is close to unity, there are some losses due to the wall effect, mostly caused by protons absorbed in the cathode and anode planes, or by 866--keV protons escaping the detector's active volume. A 3D GEANT4 simulation \cite{geant4} was performed to calculate the efficiency corrections. For the simulation, the $^{23}$Al distribution along the detector axis was extracted from the drift time of the 595--keV protons relative to the corresponding 583--keV $\gamma$--rays (see Ref. \cite{Friedman2019} for details). Unfortunately, the transverse distribution of the beam is only coarsely known, based on the relative counts in the different pads, which prevents a reliable efficiency calculation of the single pad spectrum for 583--keV protons and higher. For the combined spectrum, however, the efficiency is a weak function of the transverse beam distribution due to the 5.5-cm-diameter aperture of the cathode that limits the transverse dispersion. On the other hand, despite the excellent $\beta$ background suppression in the present experiment, extracting the intensity of the 204--keV peak from the combined spectrum still required modeling of the background with an exponential function, which was chosen based on the GEANT4 simulation, while in the pad A spectrum the 204--keV proton peak has a better signal-to-background ratio, and the background can be modeled with a linear function. For those reasons, we used both spectra to extract the 204--keV proton peak intensity as described below, while for the other proton peaks we used only the combined spectrum. The calculated efficiencies for the combined spectrum were $\epsilon = 0.98$, 0.98, 0.97 and 0.93 for the 204--, 275--, 583-- and 866--keV protons, respectively. The efficiency correction was assigned a conservative systematic uncertainty of 3\% (relative) by varying the $^{23}$Al distribution along both the detector axis and transverse plane. The diffusion of the $^{23}$Al atoms was calculated to be less than 1 cm within two lifetimes, hence the effect on the transverse distribution is contained in the systematic studies. Table \ref{tab:intensities} lists the efficiency-corrected intensities. Since the intensities of both the 583--keV and 866--keV proton peaks were extracted precisely and accurately in a region with good signal-to-background ratio in previous experiments, the agreement between the branching ratios of the 583--keV peak is an additional verification of the efficiency calculation. The extracted ratio for the 204--keV proton peak intensity is $I^{\beta p}_{204}/I^{\beta p}_{866} = 0.066\pm0.005 \textrm{ (stat.)} \pm0.002 \textrm{ (sys.)}$. An alternative approach is to use the ratio between the 204--keV and the 275--keV proton peak intensities in pad A alone. The short range of the 275--keV protons (see Fig. \ref{fig:padplane}) relative to the active volume enable reliable extraction of this ratio. We can then use the intensity ratio between the 275--keV and the 866--keV proton peaks in the combined spectrum to extract the ratio between the 204--keV and the 866--keV peaks. This approach yields $I^{\beta p}_{204}/I^{\beta p}_{866} = 0.060\pm0.005 \textrm{ (stat.)} \pm0.002 \textrm{ (sys.)}$, in agreement with the previous value. While the systematics of both values reflect the same efficiency calculation, the statistics are practically independent and the combined value is taken by averaging the above values.

\begin{table}
\centering
		\caption{Intensities of the low-energy $^{23}$Al $\beta$-delayed proton peaks from past \cite{ti95prc,pe00plb,saa11prc,Sun2015} (\cite{pe00plb} as compiled by \cite{saa11prc}) and present work, relative to the 866--keV peak intensity. Note that the 583--keV peak is contributed to by two separate decays (see Sec. \ref{sec:analysis} and Fig. \ref{fig:decay_scheme} for details).}
\begin{tabularx}{\linewidth}{p{2.8cm}p{1cm}p{1.1cm}l}
    \hline
	& \multicolumn{3}{c}{$E_{c.m.}$ (keV)}\\
	Reference                                  & 204 & 275 & 583 \\
	\hline
	Tighe \cite{ti95prc}         & 2.2(5)  & 0.9(3) & 0.7(1) \\
	Per{\"a}j{\"a}rvi \cite{pe00plb}     & 0.10(8) & 0.13(9) & 0.73(49) \\
	Saastamoinen \cite{saa11prc} & 0.34(6) & 0.45(9) & 0.69(3)  \\
	Sun \cite{Sun2015} & 0.34(12) & 0.43(15) & 0.61(12) \\
	Present Work      & 0.063(4) & 0.288(10) & 0.685(22)             \\
	\hline
\end{tabularx}
	\label{tab:intensities}
\end{table}

\begin{table*}
\centering
		\caption{Strength of the 204--keV (c.m.) $^{22}$Na(p,$\gamma$)$^{23}$Mg resonance. The strength for indirect values is calculated assuming various combinations of the branching ratios and lifetimes of the 7.79--MeV state, and 7/2 spin assignment. Upper limits are calculated within a 90\% C.L.}
\begin{tabular}{lllll}
    \hline
    method & reference & & \multicolumn{2}{c}{$\omega\gamma$ (meV)}  \\
    \hline
    direct & Stegm{\"u}ller/Seuthe \cite{se90npa,st96npa} & & \multicolumn{2}{c}{1.8(7)}  \\
    direct &	Sallaska \cite{sa10prl,sa11prc} & &  \multicolumn{2}{c}{$5.7^{+1.6}_{-0.9}$}  \\
	\\
         & & $\Gamma_p/\Gamma$ & $\tau=$ 10(3) fs \cite{je04prl} & $\tau<$ 12 fs \cite{ki16prc}   \\
	\cline{3-5}\\
	indirect & Per{\"a}j{\"a}rvi \cite{pe00plb} & 10(8)$\times10^{-3}$ & 0.4(3) & N/A \\	
	indirect & Saastamoinen \cite{saa11prc} & 3.7(9)$\times10^{-2}$ & $1.4^{+0.5}_{-0.4}$ & $>$ 0.71\\	
	indirect & Present Work & 6.5(8)$\times10^{-3}$ & 0.24(8) & $>$ 0.16\\
	\hline
\end{tabular}
	\label{tab:strengths}
\end{table*}

\section{Results and Discussion}

We measured the ratio of intensities of the 204--keV and 866--keV protons in $^{23}$Al($\beta$p)$^{22}$Na decay to be $I^{\beta p}_{204}/I^{\beta p}_{866} = 0.063(4)$. To put this value into context, it is instructive to review the previous literature on the lowest-energy proton peak from $^{23}$Al($\beta$p)$^{22}$Na decay (Table \ref{tab:intensities}).
As suggested by \cite{saa11prc}, it seems likely that the ``peak'' in Tighe \emph{et al.} \cite{ti95prc} is really misinterpreted low-energy background or noise combined with a threshold and, therefore, we believe it is reasonable to disregard the measurement of Ref. \cite{ti95prc}. The 204--keV peak intensity in Sun \emph{et al.} \cite{Sun2015} is consistent with Ref. \cite{saa11prc}, but based on very limited statistics. The present value is consistent  with the value reported by Per{\"a}j{\"a}rvi \emph{et al.} \cite{pe00plb}, which has large error bars, but the present value is a factor of 5 lower than the value in Ref. \cite{saa11prc}. The unique systematic effect in Ref. \cite{saa11prc} was related to the use of a Si implantation detector, for which a very large subtraction of background from $\beta$ particles was required to extract the proton spectrum at the lowest energies. The present work was carried out with a gas-filled detector specially designed to mitigate $\beta$-particle background in the region of interest, strongly reducing uncertainties associated with background subtraction. The gas-Si telescope used in Ref. \cite{pe00plb} was also relatively insensitive to $\beta$-particle backgrounds. Therefore, we consider the present value and the one from Ref. \cite{pe00plb} to be the most reliable. The present value provides a much smaller uncertainty that can be attributed in part to several orders-of-magnitude higher statistics.

Using the present value for the ratio $I^{\beta p}_{204}/I^{\beta p}_{866}$, we determine the absolute intensity of the 204--keV peak by adopting the absolute intensity of the 866--keV peak from Ref. \cite{saa11prc}, $I^{\beta p}_{866} = 0.41(1) \%$. The adopted value for the intensity of the 866--keV proton peak is relatively insensitive to systematic effects because of its higher energy and intensity, and it is also consistent with the absolute intensity of Ref. \cite{Sun2015}, albeit with large uncertainty. The result is $I^{\beta p}_{204} = 0.0257(17) \%$. Ref. \cite{Pollacco2013} indicated a branching ratio on the order of 0.02\%, consistent with our current value. The ratio of this value to its sum with the well-known $\beta\gamma$ intensity through the same 7.79--MeV state of $^{23}$Mg, $I^{\beta\gamma}_{7.79} =3.95(37)$\% \cite{ia06prc,Zhai2007,saa11prc} yields a proton branching ratio of $\Gamma_p/\Gamma = 6.5(8)\times10^{-3}$.

In order to calculate the resonance strength using our new value for the proton branching ratio, we must adopt a spin and lifetime for the resonance. Multiple arguments have been made for a $(7/2)^+$ spin and parity assignment, which we adopt \cite{Firestone2007,Saastamoinen2009,saa11prc}. The only finite literature value for the lifetime is $\tau =$ 10(3) fs from an in-beam gamma-ray spectroscopy measurement \cite{je04prl}. Using these values would lead to a resonance strength of $\omega\gamma = $ 0.24(8) meV, which is 7 and 23 times lower than the directly measured values from Refs. \cite{se90npa,st96npa} and \cite{sa10prl,sa11prc}, respectively (see Table \ref{tab:strengths}). A recent  lifetime measurement by Kirsebom \textit{et al.} \cite{ki16prc} did not yield a finte value, but a 95\% C.L. upper limit of $\tau < 12$ fs corresponding to a lower limit on the strength. Table \ref{tab:strengths} summarizes the resulting resonance strengths obtained using various values for the branching ratios with both lifetimes, and the direct measurement values. 

The implications of adopting various resonance strengths on $^{22}$Na production in novae was investigated through a series of 12 hydrodynamic simulations performed with the spherically symmetric, time-implicit, Lagrangian  code SHIVA \cite{jo98apj,jo16book}. The model consists of a 1.25 ONe white dwarf that accretes solar composition material at a rate of $2 \times 10^{-10}$ $M_{\odot}$ yr$^{-1}$. Material is assumed to mix with the outer layers of the white dwarf star as it piles up (50\% solar, 50\% white dwarf material). The model was coupled to a nuclear reaction network containing 118 isotopes from $^1$H to $^{48}$Ti and 630 nuclear interactions. The thermonuclear $^{22}$Na(p,$\gamma$)$^{23}$Mg reaction rate was varied based on different assumptions about the nuclear-data inputs. The total mass of ejected material was $3.8 \times 10^{28}$ g for all cases. The lowest output $^{22}$Na mass fraction of $1.1 \times 10^{-4}$ was obtained using the rates from Refs. \cite{sa10prl,sa11prc} (consistent with the value reported in \cite{sa11prc}). The highest output $^{22}$Na mass fraction of $4.3 \times 10^{-4}$ was obtained using the present branching ratio combined with the lifetime from Ref. \cite{je04prl} for the 204-keV resonance and the 275-keV resonance strength from Refs. \cite{se90npa,st96npa}. Therefore, uncertainties associated with the $^{22}$Na(p,$\gamma$)$^{23}$Mg rate translate to an almost factor-of-four variation in $^{22}$Na yields from typical ONe novae corresponding to a factor-of-two variation in maximum detectability distance for the 1.275-keV $\gamma$-ray. 

Using this new resonance strength value results in an increased $^{22}$Na production and could lead to a tension between some nova nucleosynthesis models published in the literature and observational values \cite{iy95aas,Starrfield2016} (on the contrary, see Ref. \cite{iy10ar}). However, this result relies on a single lifetime measurement with 30\% uncertainty. Shell-model calculations assuming spin and parity assignment of $(7/2)^+$ predict a much shorter lifetime of $\sim 0.6-1.7$ fs (see Ref. \cite{Jin2013} for calculation details), yielding a resonance strength of $\omega\gamma = $ 1.4-4.1 meV, which is on the same order as the direct measurements. We therefore suggest that the lifetime of the 7.79--MeV excited state of $^{23}$Mg should be re-measured, ideally to a precision better than 1 fs, prior to use with the branching ratio to obtain a resonance strength. Presently, we recommend adopting the more conservative lower limit on the resonance strength based on the present branching ratio and the lifetime limit from Ref. \cite{ki16prc}. 

\section{Summary}

In summary, we have measured the low-energy $^{23}$Al $\beta$-delayed proton intensities with the best accuracy so far, using a new detection system, GADGET \cite{Friedman2019}. The result leads to a new proton branching ratio for the key 204--keV resonance, which is a factor of 5 lower than the most precise and most recent literature value \cite{saa11prc}. If the present value is used together with the lifetime from Ref. \cite{je04prl} then the resonance strength is 7 and 23 times lower than the two direct measurements Refs. \cite{se90npa,st96npa} and \cite{sa10prl,sa11prc}, respectively, compounding the existing discrepancies in this resonance strength. Inconsistencies in nuclear data have now inflated the uncertainties in the $^{22}$Na yield from nova models by a factor of two. 

\section*{Acknowledgements}

We gratefully acknowledge the NSCL staff for collaborating on the mechanical design and fabrication of GADGET, technical assistance, and for providing the $^{23}$Al beam. We thank the NSCL gamma group for assistance with SeGA. This work was supported by the U.S. National Science Foundation under Grants No. PHY-1102511, PHY-1565546 and PHY-1913554, and the U.S. Department of Energy, Office of Science, under award No. DE-SC0016052. We also acknowledge support by the Spanish MINECO grant AYA2017--86274--P, by the E.U. FEDER funds, and by the AGAUR/Generalitat de Catalunya grant SGR-661/2017. This article benefited from discussions within the ChETEC' COST Action (CA16117).

\section*{References}
\bibliography{23Al_decay_PLB_for_submission}

\end{document}